\documentclass[11pt,twoside]{article}

\usepackage{asp2006}
\usepackage{epsf}
\usepackage{psfig}
\usepackage{lscape}

\begin{document}
\title{Relativistic inflow in the Seyfert 1 Mrk 335 revealed through X-ray absorption}   %%% Fill in title
\author{A.L.Longinotti (1), S.Sim (2), K.Nandra (3), M.Cappi (4), P. O'Neill (3)}   %%% Fill in author names
\affil{{\footnotesize (1)XMM-Newton SOC, ESAC, Madrid, Spain; (2) MPA, Garching, Germany, \\
(3) Imperial College London, (4) INAF/IASF Bologna, Italy }}    %%% Fill in author affiliations

\begin{abstract} 
The analysis of  hard X-ray features in {\it XMM-Newton} data  of the bright Sy 1 galaxy Mrk 335 is reported here. The presence of a broad, ionised iron K$\alpha$ emission line in the spectrum, first found by Gondoin et al.(2002), is confirmed. The broad line can be modeled successfully by relativistic accretion disc reflection models.
 Regardless of the underlying continuum we report, for the first time in this source, the detection of a narrow absorption feature at the rest frame energy of ~5.9 keV.  If the feature is identified with a resonance absorption line of iron in a highly ionised medium, the redshift of the line corresponds to an  inflow velocity  of ~0.11-0.15 c. 
Preliminary results from a longer (100ks) exposure are also presented.

\end{abstract}
%\vspace{-1.8cm}
\section{Results from the 30~ks archival observation}
The analysis of  the EPIC data from a  30~ks observation of the Seyfert 1 Mrk~335 (Longinotti et al. 2006) shows that the 
2-10~keV spectrum can be phenomenologically fitted with a power law, a broad  Gaussian line with E=6.22$\pm$0.16 keV, $\sigma$=0.66$\pm$0.23  keV, EW~490 eV, and a  narrow ($\sigma$ =1eV) absorption line with E=5.92$\pm$0.04 keV and EW=50~eV. These features are visible in the residuals of the spectrum plotted in Fig.1. The significance of the absorption line estimated through Monte Carlo simulations is 99.7\%.  The most obvious identification for the absorption feature is with redshifted iron K$\alpha$ resonance absorption. 
The identification with iron is favoured since the observed energy of the line is too high to be readily explained by K$\alpha$ absorption in any of the other astrophysically abundant elements. 
 We present a simple model for the inflow (centre and right panels of Fig.1), accounting approximately for relativistic and radiation pressure effects, and use Monte Carlo methods to compute synthetic spectra for qualitative comparison with the data. This modeling was developed following Sim (2005) and assuming spherically symmetric radially infalling gas. It does  show that the absorption feature can plausibly be reproduced by infalling gas providing that the feature is identified with Fe XXVI. 
 A smooth continuous flow is ruled out by the poor agreement with the data: although  the presence of a broad inverse P Cygni Fe XXVI K$\alpha$ line profile is predicted, the absorption line is insufficiently redshifted and too broad. An inflow over a limited range of radii (as discrete blobs or section of infalling gas) is more consistent with the data  (Fig.1). 
The narrowness of the absorption line tends to argue against a purely gravitational origin for the redshift of the line, but given the current data quality we stress that such an interpretation cannot be ruled out. 
\begin{figure}[h]
\begin{center}
\hbox{
\psfig{figure=longinotti_fig1.ps,height=3.5cm,width=4.5cm,angle=-90}
%\hspace{0.05cm}
\psfig{figure=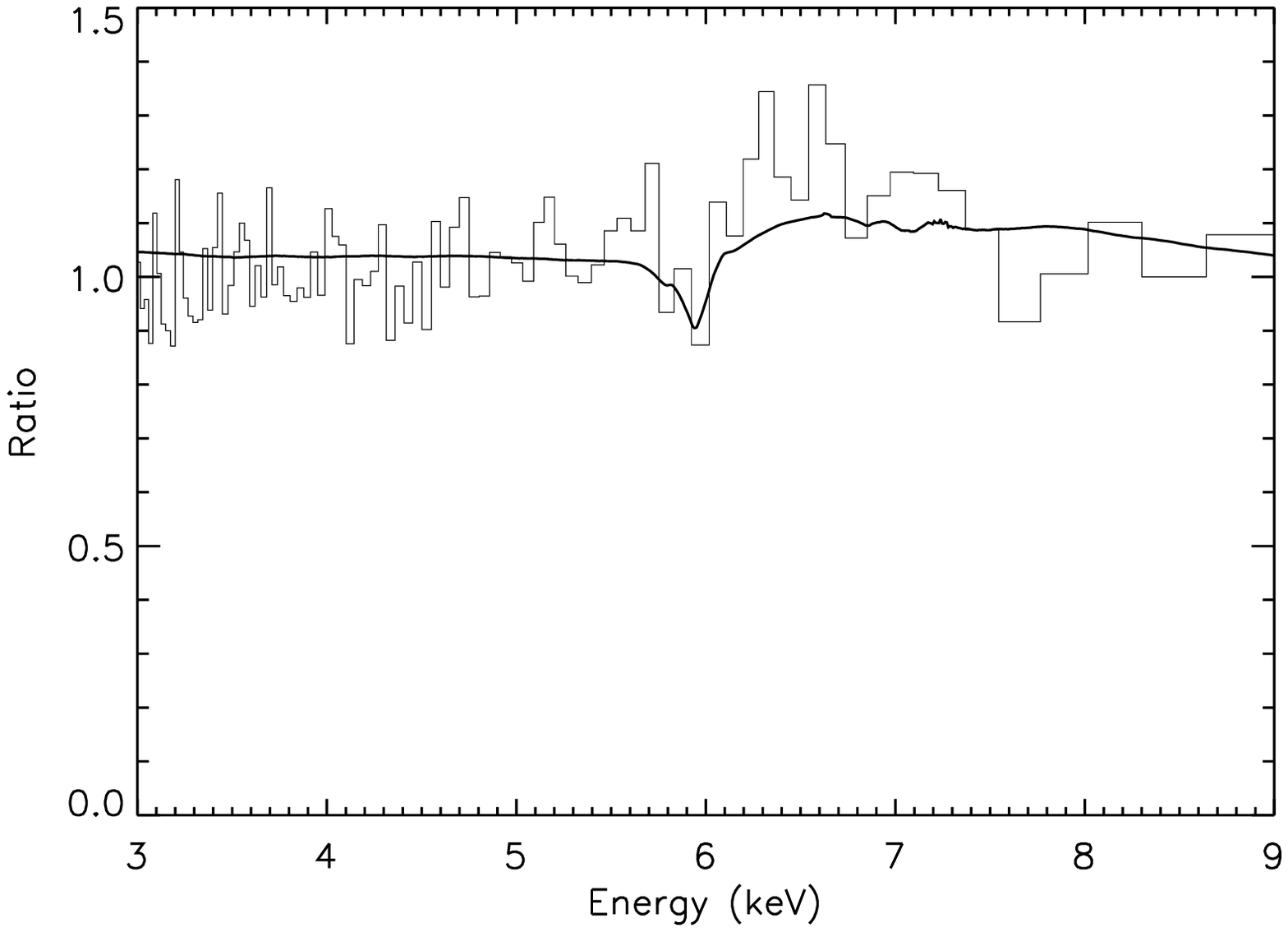,height=3.8cm,width=4.8cm}
%\hspace{0.05cm}
% un-comment the following line to include your fig1b.ps postscript file
\psfig{figure=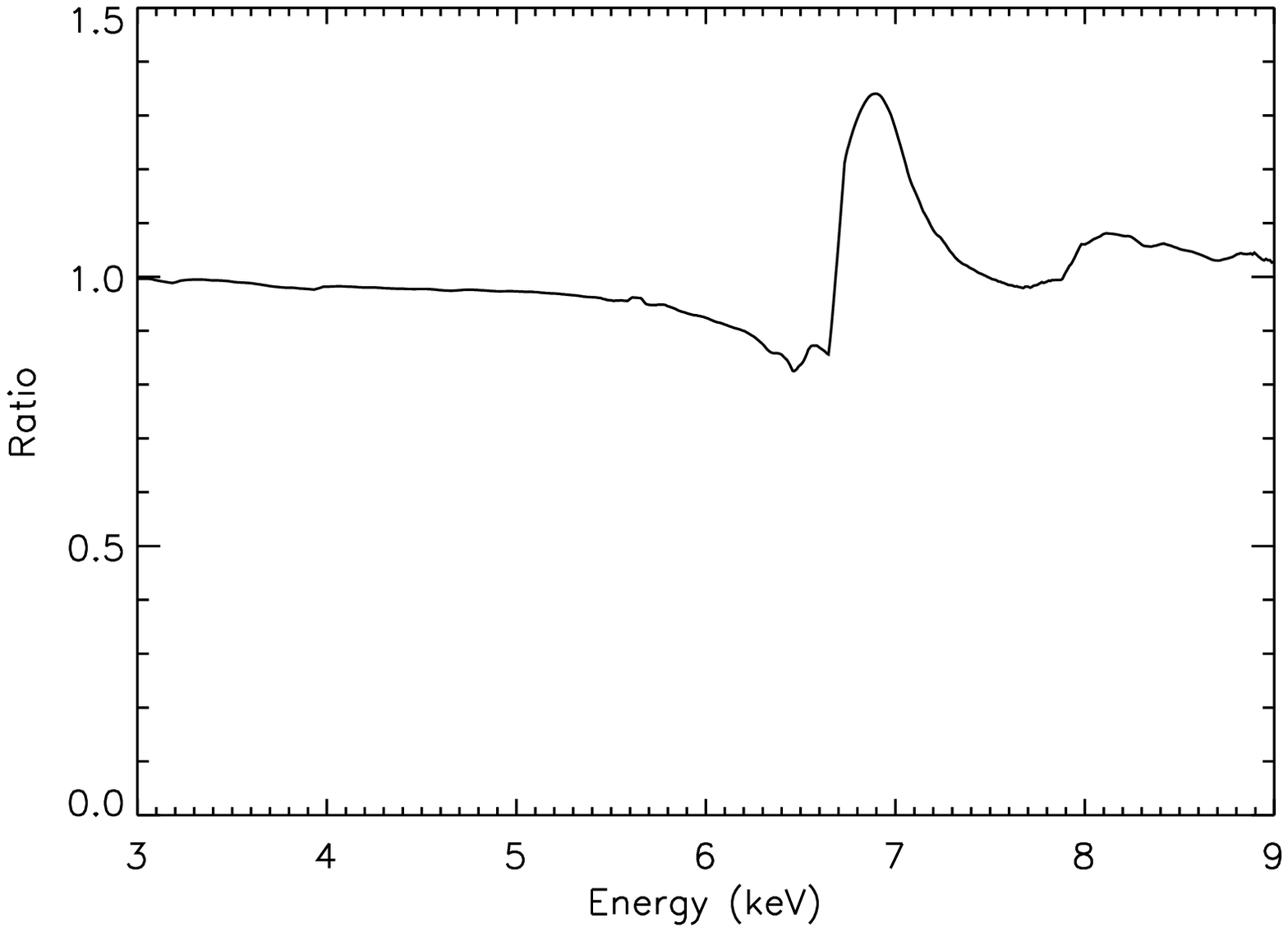,height=3.8cm,width=4.8cm}
}
\vspace{-1.2cm}
\end{center}
\caption{{\it Left}: Residuals of the EPIC spectrum of Mrk 335 fitted by a power law with $\Gamma$$\sim$2.15; {\it Centre}: Ratio of the computed 3-9~keV spectra to a power law model as produced in an inflow of gas extending from r$_{in}$=24r$_g$ to r$_{out}$=48r$_g$ (the histogram shows Mrk~335 data); {\it Right}: the same but for an inflow extending from r$_{in}$=20r$_g$ to r$_{out}$=2$\times$10$^3$r$_g$. }
\label{fig1}
\end{figure}

\section{Preliminary results from the 100~ks proprietary observation} 
\begin{figure}[h]
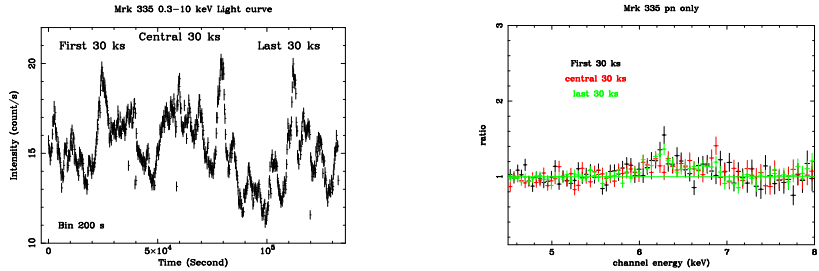

\begin{center}
\hbox{
\hspace{1cm}
\psfig{figure=longinotti_fig4.eps,height=3.5cm,width=4.5cm,angle=-90}
\hspace{1.5cm}
\psfig{figure=ratio.ps,height=3.5cm,width=4.5cm,angle=-90}
%\hspace{0.05cm}
}
\vspace{-1.2cm}
\end{center}
\caption{{\it Left}: light curve of Mrk 335; {\it Right}: ratios to a power law of the spectra extracted from the 3 portions of the total exposure. }
\end{figure}
A longer (100 ks)  XMM-Newton observation of Mrk 335 has been performed this year. The analysis of the integrated spectrum reveals a double-peaked Fe line. A broad accretion disc line is likely to be present but the peaks are due to narrower components at ~6.4 and 7 keV (O'Neill et al. in prep.).
A preliminary time-resolved analysis shows that the narrow peaks are variable. Fig. 2 shows the spectra from three portions of the light curve.
The  narrow peak at 6.4~keV, clearly present in the first 30~ks, disappears in the central portion. The 7~keV peak instead shows up only in the last 60~ks. The variable narrow lines could be tentatively associated to the presence of  flares in the light curve, but  a much more detailed and careful analysis is necessary before speculating on their origin. 
As regard to the absorption line observed in the archival 30 ks observation, it is marginally detected only in the first portion of the longer exposure. 
Further investigation on these data will hopefully clarify all the issues reported here.

%\acknowledgements %%% Text of acknowledgements runs on after this command.

%%% THE BIBLIOGRAPHY
%%%
%%% CONSULT SECTION 3 OF "INSTRUCTIONS FOR AUTHORS" FOR HOW TO USE NATBIB.
%%% AUTHORS ARE ENCOURAGED TO USE EITHER THE "THEBIBLIOGRAPY" ENVIRONMENT
%%% BY UNCOMMENTING (DELETING THE "%" SYMBOL) THE COMMANDS BELOW, OR BY
%%% USING THE BIBTEX ENVIRONMENT. TO FIND OUT WHICH IS APPLICABLE TO YOUR
%%% CONTRIBUTION, CONSULT THE VOLUME EDITORS FOR YOUR PROCEEDINGS.
%%%

\end{document}